\documentclass{article}
\usepackage{spconf,amsmath,graphicx,bm}

\title{MANDARIN TONE MODELING USING RECURRENT NEURAL NETWORKS}
\name{Hao Huang\thanks{Thanks to XYZ agency for funding.}}
\twoauthors
  {Hao Huang\sthanks{This work was supported by the National Science Fundation of China (61365005,61663044).}, Ying Hu }
	{School of Information Science and Engineering\\
	Xinjiang University\\
	Urumqi, China, 830046}
  {Haihua Xu }
	{Temasek Laboratories\\
	Nanyang Technological University\\
	Singapore, 637553}

\begin{document}
\ninept
\maketitle
\begin{abstract}
We propose an Encoder-Classifier framework to model the Mandarin tones using recurrent neural networks (RNN). In this framework, extracted frames of features for tone classification are fed in to the RNN and casted into a fixed dimensional vector (tone embedding) and then classified into tone types using a softmax layer along with other auxiliary inputs. We investigate various configurations that help to improve the model, including pooling, feature splicing and utilization of  syllable-level tone embeddings. Besides, tone embeddings and durations of the contextual syllables are exploited to facilitate tone classification. Experimental results on Mandarin tone classification show the proposed network setups improve tone classification accuracy. The results indicate that the RNN encoder-classifier based tone model flexibly accommodates heterogeneous inputs (sequential and segmental) and hence has the advantages from both the sequential classification tone models and segmental classification tone models.
\end{abstract}
\section{Introduction}
\label{sec:intro}

Tone modeling plays an important role in reducing ambiguity
in  tonal languages such as Mandarin. Utilization of tone information
has proved to be successful in improving accuracy
in Mandarin speech recognition, either by appending the pitch related tonal features with the traditional spectral features
for acoustic model training \cite{HUANG2000}, which is referred to as embedded tone modeling, or explicitly building
tone classifier and adding tone model scores in lattice re-scoring \cite{HUANG2008}.
Explicitly built tone models are also applied to detect erroneous tonal pronunciations in computer-assisted language learning (CALL) \cite{CHENG2012}. Compared with embedded tone modeling,  the explicit tone modeling approach is capable of exploiting the
supra-segmental nature of the tones or using better tone classifier to improve system performance.

There are two major approaches to explicit tone modeling:  sequence based tone modeling and segment based tone modeling. Sequence based tone models accept sequential observations while the segment based tone models use fixed dimension feature vector. Because articulation of human is temporal and output of  pitch-related feature extraction processing  is frame based, to model the tones using sequential model is natural and reasonable. The  sequenced model can be hidden Markov model (HMM) \cite{WANG1997} or hidden conditional random fields (HCRFs) \cite{Gunawardana2005}, etc. The shortcoming of this modeling method is that it is difficult for the sequence based models to utilize  segment based information from the contextual tones.
For example, great efforts are needed to consider the pitch related features of contextual syllable \cite{HUANG2007}. An alternative to sequence based method is to classify the tones using segmental classifiers such as Gaussian mixture model (GMM) \cite{QIAN2003}, support vector machine (SVM) \cite{Peng2005}  and neural network (NN) \cite{Chen1995}. Besides,  stochastic polynomial trajectory model (SPTM) \cite{Cao2004}, and decision tree based tone classifier \cite{Wong2004} have also been tried. Though the proposed methods can be effective, there has to be manually conversion of the temporal observation sequence into fixed dimensional features, which requires expertise and experience and might not get the optimal results. 

% Below is an example of how to insert images. Delete the ``\vspace'' line,
% uncomment the preceding line ``\centerline...'' and replace ``imageX.ps''
% with a  suitable PostScript file name.
% -------------------------------------------------------------------------
\begin{figure*}[htb]
	
	\begin{minipage}[b]{1.0\linewidth}
		\centering
		\centerline{\includegraphics[width=16.5cm]{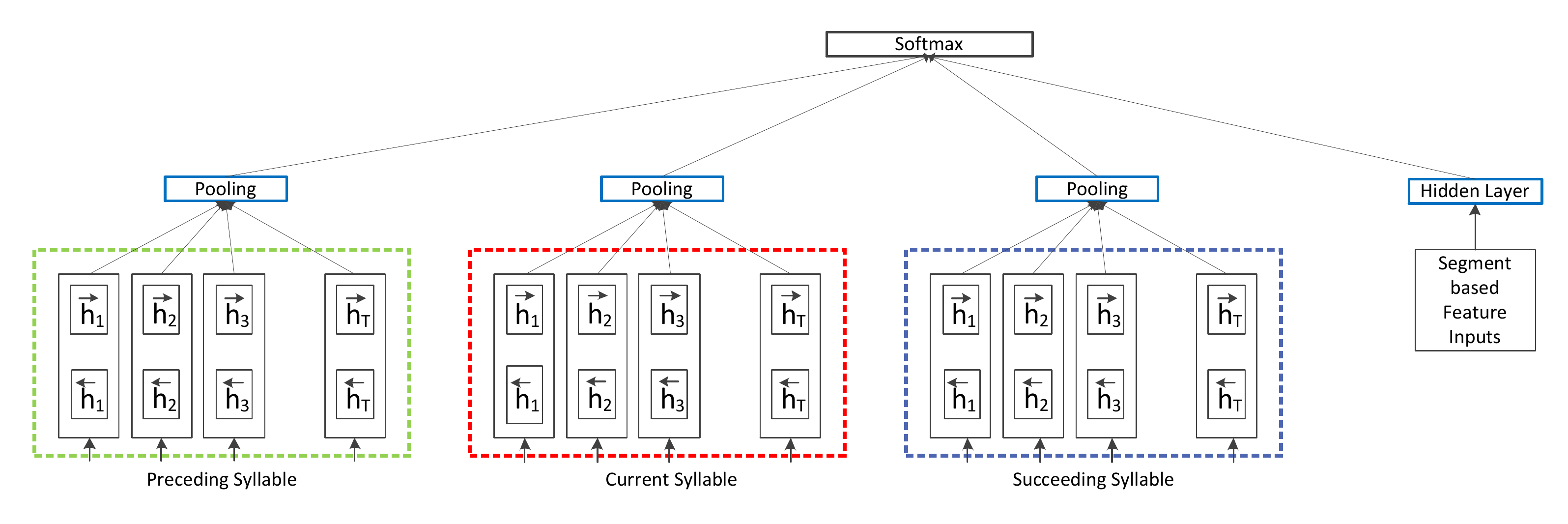}}
	\end{minipage}
	%	%
	%	\begin{minipage}[b]{.48\linewidth}
	%		\centering
	%		\centerline{\includegraphics[width=4.0cm]{image3}}
	%		%  \vspace{1.5cm}
	%		\centerline{(b) Results 3}\medskip
	%	\end{minipage}
	%	\hfill
	%	\begin{minipage}[b]{0.48\linewidth}
	%		\centering
	%		\centerline{\includegraphics[width=4.0cm]{image4}}
	%		%  \vspace{1.5cm}
	%		\centerline{(c) Result 4}\medskip
	%	\end{minipage}
	%	%
	\caption{Tone modeling using RNN encoder-classifier based framework, including an auxiliary input layer}
	\label{fig:res}
\end{figure*}

Recently, the progress in deep learning  has shown great success in
various applications such as computer vision
\cite{KRIZHEVSKY2012} and speech
recognition \cite{DAHL2012,HINTON2012}. Moreover, many recent works showed that neural
networks can be successfully used in a number
of tasks in natural language processing (NLP), such as machine translation \cite{CHO2014}, word embedding extraction \cite{MIKOLOV2013} and sentence classification \cite{LEE2016}. For tone modeling, authors in \cite{CHEN2014} proposed to use DNN as tone classifier with manually designed tone features and obtained better results with traditional classifier such SVM, which belongs to the segmental classifier based framework. Authors in \cite{RYANT2014} use DNN for
frame-level tone classification using co-articulation features extracted from raw MFCCs as input without pitch tracking.  \cite{CHEN2016}  proposed a method that
fully automates tone classification of syllables in Mandarin Chinese,
which takes as input the raw tone data and uses convolutional
neural networks (CNN) to classify the tones using raw MFCC other than manually edited F0 (fundamental frequency).

In NLP research community, RNN is often used to convert an input word sequence into a fixed dimensional vector, such as in the Encoder-Decoder framework in machine translation \cite{CHO2014} and distributed sentence representation \cite{KIROS2015}. For sentence classification task, authors in \cite{LEE2016} in proposed to  map the entire word sequence into a vector for classification. Concerning the possible heterogeneous inputs (sequential or segmental) of the tone model, it is appealing to convert pitch related sequence into a fixed-dimensional vector and classify into tone types along with other arbitrary segmental input features. Inspired by this, we proposed to use RNN as a tone modeling framework. We convert the frame based pitch-related observation sequence into a fixed dimensional vector and classified into tone types using a softmax layer. We explore to use methods such as feature splicing, mean pooling and full-syllable tone embedding to improve tone classification accuracy. Also, tone embeddings and durations of contextual syllables are utilized to facilitate tone classification.  According to the results, we found the Encoder-Classifier tone model based on RNN combines the advantages of both the sequential tone model and segment classification model. The framework is convenient to model long span features, transfer sequential observations to fixed dimensional vector,  automate the extraction of tone nuclei of the tones, and easy to incorporate contextual tonal information, hence to combine with other segment based features. We also show a preliminary comparison with the ML and MMI trained GMM-HMM tone models.

The remainder of this paper is organized as follows: In Section
2, the Encoder-Classifier based tone model is described.
Section 3 gives the experiments and the results along with various experimental setups.
Finally in section 4 the conclusions drawn from the work
are given and the future work is presented.

\section{RNN Encoder-Classifier Tone modeling}
Here, we briefly describe the underlying framework, called RNN Encoder-Classifier. The framework comprises two parts: The first part maps the input observation sequence to a fixed dimensional vector (tone embedding) using the RNN. The second part classifies the tone embedding (or a few tone embeddings of successive syllables) into the five tone labels. The framework enables the network to be trained to discriminate between tones from variable-length speech segments and other manually designed segmental features can be conveniently incorporated as well. The framework is illustrated in Figure 1.

In the Encoder-Classifier framework, an encoder reads the input observation sequence of vectors ${\mathbf x} = ({\mathbf x}_1, \cdots , {\mathbf x}_T )$ and converts them into a compressed vector $\mathbf c$. The most common approach is to use an RNN, which is a neural network that consists of a hidden state $\mathbf h$ an output $y$ which operates on the variable length
sequence ${\mathbf x} $. At each time $t$, the hidden state ${\mathbf h}_t$ of the RNN is update by
\begin{equation}
{\mathbf h}_t=f({\mathbf x}_t,{\mathbf h}_{t-1}).
\label{E201710211121}
\end{equation}
The compressed vector $\mathbf c$ is a function of the output sequence of the hidden states:
\begin{equation}
{\mathbf c}=q({\mathbf h}_1,\cdots,{\mathbf h}_{T}).
\end{equation}
The layer for this purpose is referred to as the pooling layer.  $f$ and $q$ are some nonlinear functions.   Given the vector $\mathbf c$, the predict layer define a probabilities over the tone labels $\mathbf y$ by
\begin{equation}
{p({\mathbf y}|{\mathbf x}})={g}({\mathbf c}),
\label{E201710201506}
\end{equation} 
where ${\mathbf y}=(y_1,\cdots,y_5)$ represent the five tone labels given $\mathbf x$ and $g$ is a nonlinear, potentially multi-layered, function that outputs the  posterior probabilities of $\mathbf y$. 
The two components of the proposed RNN Encoder-Classifier are jointly trained to maximize the conditional log-likelihood 
\begin{equation}
\mathop{\max}_{\theta} \frac{1}{N} \log p_{\bm \theta}({\tilde y}_n|{\mathbf x}_n),
\end{equation}
where $\bm \theta$ is the set of the model parameters and $({\mathbf x}_n,{\tilde 
	y}_n) $ is the pair from the training set.
The more sophisticated network structures that improve the model will be described in the experiments.

\section{Experiments and Results}
\label{sec:pagestyle}
\subsection{The database}

The proposed method is evaluated on a Mandarin speech recognition database. The '863 project' Mandarin speech database is used, which is a corpus for continuous speech recognition collected for the  Chinese National '863 Project'. It has a total of about 110-hour recordings spoken by 160 speakers (80 females and 80 males). The database contains 92,243 utterances. In our experiments, 86,271 utterances are selected as the training set, the rest are used for evaluation. Table 1 lists the two subsets of the database and the distribution of the tones in the corpus.

The pitch related features were extracted by using the {\ttfamily compute 
	-and-process-kaldi-pitch-feats} command in the Kaldi speech recognition toolkit \cite{POVEY2011} with default command options. Each input observation vector has 3 dimensions: The log-pitch with Probability of Voicing (POV) weighted mean subtraction, and the time derivative of log-pitch and the warped Normalized Cross Correlation Function (NCCF) \cite{Ghahremani2014}.  The features are normalized by speaker dependent means and variances. Before training the tone model, we use DNN-HMM based acoustic models trained for speech recognition to obtain phone boundary information for each phone-segment using  Viterbi alignment.

\subsection{Results}
\subsubsection{Baseline}
For the baseline, the inputs to RNN are only the observations of the final portion of the syllable, which conforms to the common knowledge that tone is perceived mainly in the final part of a syllable. The input observation uses current frame (without feature splicing). As for the RNN, 
Elman neural network is used, which keeps track of the previous hidden layer
states through its recurrent connections:
\begin{equation}
{\mathbf h}_t={\sigma}({\mathbf W}{\mathbf x}_t+{\mathbf V}{\mathbf h}_{t-1}+{\mathbf b}),
\label{E201710211120}
\end{equation}
 where $\mathbf W$ and  $\mathbf V$ are the weight matrices and $\mathbf b$ is the bias vector. $\sigma(\cdot)$ is the sigmoid function. 
 The dimension of the input observations $\mathbf x$ is 3. The hidden layer size is set to 250.  As for the pooling layer, which combines the sequence of vectors $\mathbf h$ into a single vector $\mathbf c$ that represent the tone, we first experiment with {\it Last Pooling},  which takes the hidden layer output of the last frame as the tone embedding:
\begin{equation}
{\mathbf c}={\mathbf h}_T.
\end{equation}
The output of the pooling layer $\mathbf c$ is fed into the classifier, for which we
use a simple softmax function as $g$ in \eqref{E201710201506} such that:
\begin{equation}
{p({\mathbf y}|{\mathbf x}})={{\rm softmax}}({\mathbf c}).
\end{equation}
Using the above setups, the baseline tone classification accuracy on the test set is 70.7\%.

\begin{table}
	\renewcommand\arraystretch{1.1}
	\caption{Number of samples of the tones in the speech corpus (K)}
	\centering
	%\begin{tabular}{p{62pt} |p{50pt}  p{35pt} p{40pt} p{40pt}}
	\begin{tabular}{|p{20pt} |p{23pt} p{23pt} p{23pt}  p{23pt} p{23pt} p{23pt}|}		
		\hline
		Data & \hfil Tone0 &\hfil Tone1& \hfil Tone2 
		& \hfil Tone3 & \hfil  Tone4 &\hfil  Total\\
		\hline
		\hline
		
		Train     &\hfil 60.4  &\hfil 219.9 &\hfil 234.7 &\hfil 172.5 &\hfil 359.7 & 1047.3\\
		Test     &\hfil 4.1  &\hfil  15.1 &\hfil 16.2 &\hfil 11.4 &\hfil 24.6 & \hfil 73.4\\		
		\hline
	\end{tabular}
\end{table}

\subsubsection{Feature splicing}
Feature splicing has shown to be effective in speech recognition. For GMM-HMM based acoustic model, utilization of spliced features is more complicated and is often achieved by using  LDA-MLLT (Linear Discriminant Analysis and Maximum Likelihood Linear Transform \cite{SAON2000}) method. In comparison, DNN/RNN based acoustic model splicing the input features is straight-forward to implement.
In this setup, each feature vector is extracted from a small overlapping window of observation frames. The input feature of the RNN has a dimension of 27 by concatenating a context of 4 frames left and right
of the current frame (9 frames in total). It is shown in Table 2 that feature splicing has greatly improves tone classification performance, which yields 5.3\% absolute improvement compare with the baseline using single frame input.

\subsubsection{Pooling}
%\paragraph{Average Pooling}

%\indent{\it A. Average Pooling} 
In RNN based tone modeling, the hidden layer outputs the dynamic state of the pitch contours. In baseline RNN with last pooling, using the hidden state of the last frame does not fully consider the  pitch contour of the entire segment. We tried another two pooling mechanisms: {\it Average pooling} and {\it Max pooling}. Average pooling compute tone embedding by averaging all the hidden vectors
 \begin{equation}
 {\mathbf c} ={\mathbf h}_{avg}=\frac{1}{T}\sum_t^T {\mathbf h}_t,
 \end{equation}
and max pooling takes the element wise maximum of ${\mathbf h}_t$
 \begin{equation}
{\mathbf c} ={\mathbf h}_{max}.
\end{equation}
It is found with average pooling,  tone classification accuracy improves by 1.3\% absolute over the baseline with spliced input features (76.0\%).  Max pooling shows a comparable performance (77.4\%). We think the last hidden state ${\mathbf h}_T$ can not long span tone information over the whole segment. Utilization of  max pooling or average pooling will consider pitch trends in the entire segment.
\begin{table}[t]
	
	\begin{minipage}[b]{1.0\linewidth}
		\centering
		\caption{\label{table1} {Tone classification accuracy of the RNN based  models (\%)}}
		\begin{tabular}{|p{45pt}|p{85pt}|p{28pt}p{28pt}|}
			\hline
			Input &{Configuration}			 & \hfil Train & \hfil Test \\
			\hline
			\hline
			
			&Baseline				& \hfil 80.9  	&\hfil 70.7\\
Final 			&Splicing				& \hfil 83.8  	&\hfil 76.0 \\				
Observations			&Average Pooling				& \hfil 85.9  	&\hfil 77.3\\	
			&Max Pooling				& \hfil 85.8  	&\hfil 77.4\\	
			\hline
			&Average Pooling	\textcircled{1}			& \hfil 88.0  	&\hfil 79.7\\	
			Syllable &Max Pooling				& \hfil 88.3  	&\hfil 79.3\\				
Observations			&Backward RNN			& \hfil 88.1    	&\hfil 80.2\\
			&Bi-directional RNN 	& \hfil 88.2    	&\hfil 79.4\\
			\hline
			&\textcircled{1}	+Preceeding  				& \hfil 88.7  	&\hfil 80.6\\	
Syllable			&\textcircled{1}	+Succeeding 				& \hfil 88.9 	&\hfil 80.9\\
Observations			&\textcircled{1}	+Both				& \hfil  89.7	&\hfil 82.1\\
			&\textcircled{1}	+Duration				& \hfil  90.4	&\hfil 82.9 \\
			%+Attention				& \hfil & \hfil \\							
			\hline
		\end{tabular}
	\end{minipage}
	
\end{table}

\subsubsection{Forward, Backward and Bidirectional Variants}

The baseline RNN described in \eqref{E201710211121} and \eqref{E201710211120} reads the input sequence ${\mathbf x}=({\mathbf x}_1,\cdots,{\mathbf x}_T)$ in order starting from the first
frame ${\mathbf x}_1$ to the last one ${\mathbf x}_T$ .
It is also possible to take into account
future information with a single backward pass. A more
appealing model would consider both past and future information
at the same time. 
Researchers have propose to use a bidirectional RNN,
% (BiRNN, Schuster and Paliwal, 1997), which has been
successfully used recently in speech recognition. 
%(see, e.g., Graves et al., 2013). 
%Based on the above discoveries, 
It could be interesting to see whether BiRNN is helpful to Encoder-Classifier based tone model.

A BiRNN consists of a forward and a backward RNNs. The forward RNN $\overrightarrow{f}$ reads the input sequence as it is ordered (from ${\mathbf x}_1$ to the last one ${\mathbf x}_{T}$) and calculates a sequence of forward hidden states $(\overrightarrow{{\mathbf h}_1},\cdots,\overrightarrow{{\mathbf h}_T})$. The backward RNN  $\overleftarrow{f}$ reads the sequence in the reverse order  (from ${\mathbf x}_T$ to ${\mathbf x}_{1}$), resulting in a sequence of backward hidden states $(\overleftarrow{{\mathbf h}_T},\cdots,\overleftarrow{{\mathbf h}_1})$. In Encoder-Classifier based tone modeling using backward RNN, the tone embedding is the averaged backward hidden states
\begin{equation}
{\mathbf c} =\overleftarrow{{\mathbf h}}_{avg} =\frac{1}{T}\sum_t^T \overleftarrow{{\mathbf h}_t}. 
\end{equation}
When BiRNN is used, the tone embedding is computed by concatenating the averaged forward hidden states and the  averaged backward hidden states:
\begin{equation}
 {\mathbf c} =\overrightarrow{{\mathbf h}}_{avg}\oplus \overleftarrow{{\mathbf h}}_{avg}, %=\frac{1}{T}\sum_t^T {\mathbf h}_t,
\end{equation}
where $\oplus$ denotes the concatenation operation. 
The hidden layer of the backward RNN also has 250 hidden units. It is shown in Table 2 that the backward RNN shows slight better accuracy on the test set, but the Bi-directional RNN shows even slight lower accuracy. We think the tone classification model, which classify the entire model into a class label does not benefit much from the BiRNN. This task is different from traditional sequence labeling task, which uses forward and backward information to help predict current frame label. For segmental classification task, the output of average pooling has to some extent contained the overall contour information for classifying the whole input segment.

\subsubsection{Using Full Syllable Tone Embedding}
In traditional tone classification task, it is commonly assumed the pitch values exists in the voiced part of a syllable. In the baseline, tone models classification are conducted on features extracted from the final part of a syllable. Here we train the RNN using features spanning over the whole syllable. It is shown substantial  improvement (2.6\% absolute) over the result from using only the pitch values extracted from the final portion (77.3\%). This indicates that the boundaries obtained by the DNN speech recognizer might be inaccurate or they might not the  best for tone classification task. Another reason is that features for tone discrimination do not merely exist in the final part of syllable. As stated in \cite{WANG2008}, a syllable F0 contour could be divided into three segments: onset course, tone nucleus and offset course. Tone nucleus is a piece of F0 contour that represents pitch targets of the lexical tone, which contains the most critical information for tonality perception. Tone nucleus obtained within the final part of a syllable might be lost. The tone embeddings extracted by RNN using full-syllable observations is capable of detecting the underlying crucial information to discriminate the tones.

\subsubsection{Integration of Supra-Segmental Information}
As claimed, the superiority of Encoder-Classifier based tone model to the frame-synchronous model (HMM or HCRFs)  is  its convenience of integrating supplemental segment-based features. To achieve this goal, we extend the model by adding auxiliary inputs
which provides complementary information
to the input layer. The auxiliary input layer can be used to feed in arbitrary
additional information, either automatically learned features or manually designed features.

\

 \noindent{\it A. Contextual Tone Embeddings}

Tone classification on continuous speech is much more difficult than isolated tone classification task because of the co-articulation effect, that is, tone is not only determined by pitch contour of the current syllable, but also heavily influenced by the behavior of left and right tones. How to characterize tone variation with different contexts has been widely discussed. The context-dependent HMMs \cite{WANG1997} were selected by observing the co-articulation effect of neighboring tones were investigated. The decision tree-based clustering method was applied to obtain the optimal context-dependent models \cite{Cao2004}. These approaches showed effectiveness in reducing diversity of tone with different contexts; however, they did not really integrate context tone features into either training or recognition.

It is straightforward to model the contextual tone information for segment based models. In \cite{QIAN2003}, overlapped ditone model is proposed which integrates contextual pitch features for GMM based tone models.  However, it needs manually designed features.  For pure frame based model such HMM or CRFs, modeling the context feature is rather difficult. For example, to utilize the supra-segmental nature of Mandarin tones, we proposed a feature extraction method for HMM based tone modeling in \cite{HUANG2007}. The method uses linear transforms to project F0 features of neighboring syllables as compensations, and adds them to the original F0 features of the current syllable. The transforms are discriminatively trained by using an minimum Beyesian risk (MBR) objective function. Though the method is successful, still it needs manual time-normalization of the pitch values of preceding syllable and following syllable. The Encoder-Classifier based model makes the use of the supra-segmental features easy to implement. To use the contextual tone features, the input of softmax classifier is
\begin{equation}
{\mathbf c}={\mathbf h}^{prec}_{avg} \oplus {\mathbf h}^{curr}_{avg} \oplus  {\mathbf h}^{succ}_{avg},
\end{equation}
where ${\mathbf h}^{prec}_{avg}$, ${\mathbf h}^{curr}_{avg}$ and ${\mathbf h}^{succ}_{avg}$ are respectively the average hidden vectors of the preceding syllable, the current syllable and the succeeding syllable RNN. 

\

\noindent{\it B. Duration}

Utilization of duration as feature has shown to be effective in tone classification. The Encoder-Classifier framework makes it easy to use duration feature along with the sequential input observations. We first normalize the durations (in frames) of each segment by the mean and variance of all the segments. Then the durations
of the successive 3 three syllables (current syllable, preceding syllable and succeeding 
syllable)  are used to form a 3-dim duration feature vector $\mathbf {d}$. The duration vector is sent into a sigmoid hidden layer (with the size of 10) and the outputs are concatenated with the tone embeddings of the contiguous three syllable as the total input of the final softmax layer,
\begin{equation}
{\mathbf c}={\mathbf h}^{prec}_{avg} \oplus {\mathbf h}^{curr}_{avg} \oplus  {\mathbf h}^{succ}_{avg} \oplus \sigma({\mathbf {d}}).
\end{equation}
It is shown in Table 2 that combination of the tone embeddings from successive syllables (preceding, succeeding or both) shows better results. Tone classification accuracy using tone embeddings from three successive syllables is improved by 2.4\% absolute better than that using tone embedding of only the current syllable (79.7\% on the test set). By adding the duration feature, a further slight improvement (0.8\%) is obtained. These results indicate the proposed model accommodates both automatically learned feature and arbitrary manually designed features.

%\begin{table}[t]
%	
%	\begin{minipage}[b]{1.0\linewidth}
%		\centering
%		\caption{\label{table1} {Results of different feature splicing setups}}
%		\begin{tabular}{|p{55pt}|p{25pt}p{25pt}|}
%			\hline
%			{~~~Context}			 & \hfil Train & \hfil Test \\
%			\hline
%			\hline
%			~~~CXT=1				& \hfil  0.375	&\hfil 0\\
%			~~~CXT=3				& \hfil  0.375	&\frac{num}{den}\hfil 0\\
%			~~~CXT=5				& \hfil  0.375	&\hfil 0\\
%			~~~CXT=7				& \hfil  0.375	&\hfil 0\\
%			~~~CXT=9				& \hfil  0.375	&\hfil 0\\ 
%			~~~CXT=11				& \hfil  0.375	&\hfil 0\\															
%			\hline
%		\end{tabular}
%	\end{minipage}
%	
%\end{table}

\subsubsection{GMM-HMM based results}
We show a preliminary comparison results with the GMM-HMM based tone model. For GMM-HMM based tone modeling, we trained context-independent GMM-HMMs using ML estimation and the run MMI discriminative training for further improvement. The classification accuracy on the test set is 70.4\% for ML trained GMM-HMMs and improves to 75.7\% for MMI trained GMM-HMMs. By comparing the results from RNN in Table 2 and those from GMM-HMMs in Table 3, we have seen the superiority of the RNN based tone models to the GMM-HMMs base tone models.

\begin{table}[t]
	
	\begin{minipage}[b]{1.0\linewidth}
		\centering
		\caption{\label{table1} {Results on GMM-HMM based tone models}}
		\begin{tabular}{|p{35pt}|p{26pt}p{25pt}|}
			\hline
			{Criterion}			 & \hfil Train & \hfil Test \\
			\hline
			\hline
			MLE				& \hfil 71.5 	&\hfil 70.4\\
			MMI				& \hfil 76.3 	&\hfil 75.7\\

			\hline
		\end{tabular}
	\end{minipage}
	
\end{table}

%\vspace{-2.5mm}

\section{Conclusion and Future Work}
\label{sec:typestyle}

We have proposed an Encoder-Classifier tone modeling framework using RNN for Mandarin tone classification task.
In the framework, extracted frames of features for tone recognition are fed in to the RNN and casted into a fixed dimensional vector (tone embeddings) and classified into tones using a softmax classifier. We show various method to improve the model. The proposed Encoder-Classifier framework is promising in that it allows both sequence based observations and arbitrarily designed segment based features.

As discussed, the presented framework is at the early stage. Methods can be used to for further improvement of the model, such as context-dependent model and attention mechanism.  A comprehensive experimental study needs to be carried out. Thorough comparisons with DNN-HMM based tone model and  segment based models (SVM and DNN) should be  investigated. End-to-end learning for the tone models without explicit pitch feature extraction could be further explored. Its application to tonal mispronunciation detection task for computer assisted language learning is remained and to be explored.

%\vfill
%\pagebreak

\vfill\pagebreak
\bibliographystyle{IEEEbib}

\begin{thebibliography}{99}

\bibitem{HUANG2000}
C. H. Huang, and F. Side, ``Pitch tracking and tone features for Mandarin
speech recognition," in {\it Proc. of ICASSP}, 1523-1526, 2000.

\bibitem{HUANG2008} H. Huang, and J. Zhu, ``Discriminative incorporation of explicitly trained tone models into lattice based rescoring for Mandarin speech recognition,"c in {\it Proc. of ICASSP}, 1541-1544, 2008.

\bibitem{CHENG2012} J. Cheng, ``Automatic tone assessment of non-native Mandarin speakers," in {\it Proc. Interspeech}, 1299-1302, 2012.
	
	




\bibitem{WANG1997} H. M. Wang, T. H. Ho, and R. C. Yang. ``Complete recognition of continuous Mandarin speech for Chinese language with very large vocabulary but limited training data," IEEE Transactions on Speech and Audio Processing, 5(2): 196-201, 1997.

	
\bibitem{Gunawardana2005} A. Gunawardana, M. Hahajan, A. Acero, and J. C. Platt, ``Hidden conditional random fields for phone
classification," In {Proc. of Eurospeech}, 1117-1120, 2005.

\bibitem{QIAN2003} Y. Qian, T. Lee, and J. Y. Li. ``Overlapped ditone modeling for tone recognition in continuous
Cantonese speech," in {\it  Proc. of Eurospeech}, 1845-1848, 2003.


\bibitem{HUANG2007} H. Huang, J. Zhu, ``Discriminative tonal feature extraction method in mandarin speech recognition," The Journal of China Universities of Posts and Telecommunications, 14(4):126-130, 2007.

\bibitem{Peng2005} G. Peng, and W. S. Wang, ``Tone recognition of continuous Cantonese speech based on support vector
machines. Speech Communication, 45, 49-62, 2005.



\bibitem{Chen1995} S. H. Chen, Y. R. Wang, ``Tone Recognition of Continuous Mandarin Speech Based on Neural Networks," {IEEE Trans. on Speech and Audio Processing}, vol. 3(2), 146-150, 1995.



\bibitem{Thubthong2001} N. Thubthong, and B. Kijsirikul. ``Tone recognition of continuous Thai speech under tonal assimilation and
declination effects using half-tone model," International Journal of Uncertainty, Fuzziness and Knowledge-Based
Systems, 9(6), 815-825, 2001.

\bibitem{Cao2004} Y. Cao, S. W. Zhang, T. Y. Huang, ``Tone modeling for continuous Mandarin speech recognition,"
{International Journal of Speech Technology}, 7(2-3): 115-128, 2004.


\bibitem{Wong2004} P. F. Wong, and M. H. Siu, ``Decision tree based tone modeling for Chinese speech recognition," in
{Proc. of ICASSP}, 905-908, 2004.



\bibitem{KRIZHEVSKY2012} A. Krizhevsky, I.
Sutskever, and G. Hinton, ``ImageNet
classification with deep convolutional neural
networks," In Advances in Neural Information
Processing Systems (NIPS), 2012.


\bibitem{DAHL2012} G. E. Dahl, D. Yu, L. Deng,
and A. Acero. ``Context-dependent pretrained
deep neural networks for large vocabulary
speech recognition," IEEE Transactions on Audio,
Speech, and Language Processing, 20(1):33-42, 2012.


\bibitem{HINTON2012} G. Hinton, L. Deng, D. Yu, G. Dahl, A. R. Mohamed, N. Jaitly, A. Senior, V. Vanhoucke, P. Nguyen and T. Sainath, and B. Kingsbury, “Deep neural networks for acoustic modeling in speech recognition: The shared views of four research groups,” IEEE Signal Process. Mag. 29, 82–97, (2012).


\bibitem{CHO2014} K. Cho, B. Merrienboer, C. Gulcehre, F. Bougares, H. Schwenk, and Y. Bengio, ``Learning phrase representations using RNN encoder-decoder for statistical machine translation," In Arxiv preprint arXiv:1406.1078,
2014.

\bibitem{MIKOLOV2013} T. Mikolov, I. Sutskever,
K. Chen, G. Corrado, and J. Dean. ``Distributed
representations of words and phrases and
their compositionality," In Advances in Neural Information
Processing Systems, 3111-3119, 2013.


\bibitem{LEE2016} Ji, Young Lee, and F. Dernoncourt. "Sequential Short-Text Classification with Recurrent and Convolutional Neural Networks." In {Proc. of NAACL-HLT}, 515–520,
515-520, 2016.


\bibitem{CHEN2014} M. Chen, Z. Yang Z, W. Liu, ``Deep neural networks for Mandarin tone recognition," in  Proc. of International Joint Conference on Neural Networks (IJCNN), 1154-1158, 2014.

\bibitem{RYANT2014}
N. Ryant, M. Slaney, M. Liberman, E. Shriberg, and J. Yuan,
“Highly accurate Mandarin tone classification in the absence of
pitch information,” in Proceedings of Speech Prosody, vol. 7,
2014.

\bibitem{CHEN2016} C. Chen, R. Bunescu, L. Xu, C. Liu, ``Tone Classification in Mandarin Chinese Using Convolutional Neural Networks. in {\it Proc. of Interspeech}, 2016.

\bibitem{KIROS2015} R. Kiros, Y. Zhu, R. Salakhutdinov. Skip-thought vectors. Advances in Neural
Information Processing Systems (NIPS): 3276-3284, 2015.
	%Type = Inproceedings
\bibitem{POVEY2011}
D. Povey, {A. Ghoshal},
{G. Boulianne}, {L. Burget},
{O. Glembek},{N. Goel},
{M. Hannemann}, {P. Motl\'{i}\v{c}ek},
{Y. Qian}, {P. Schwarz},
{J. Silovsk\'y}, {G. Stemmer}, and
{K. Vesely},
{``The Kaldi speech recognition toolkit,"} in:
{{ASRU}}, {7304--7308}, (2011).


\bibitem {Ghahremani2014} P. Ghahremani, B. BabaAli, D. Povey, K. Riedhammer, J. Trmal, and S. Khudanpur, ``A Pitch Extraction Algorithm Tuned for Automatic Speech Recognition," in {\it Proc. of ICASSP,} 2014.




\bibitem{SAON2000} G. Saon, M. Padmanabhan, R. Gopinath and S. Chen, ``Maximum likelihood discriminant feature
spaces," in {\it Proc. ICASSP }, vol. II, pp. 1129-1132, 2000.




\bibitem{WANG2008} X. D. Wang, K. Hirose, and J. S. Zhang, ``Tone Recognition of Continuous Mandari n Speech Based on Tone Nucleus Model and Neural Network," IEICE Transactions on Information and Systems, E91-D(6):1748-1755, 2008.




% -------------------------------------------------------------------------

%\bibliography{strings,refs}
\end{thebibliography}

\end{document}